\documentclass[11pt,twoside]{article}
\usepackage{asp2004}
\usepackage{psfig}
\usepackage{epsf}
\usepackage{graphics}
\usepackage{lscape}
\def\edcomment#1{\iffalse\marginpar{\raggedright\sl#1\/}\else\relax\fi}
\reversemarginpar

\begin{document}
\title{On the Correlation of IR and Opticical Variability in NGC4151}
\author{V. L. Oknyanskij, V.\ M.\ Lyuty,  O.\ G.\ Taranova, V.\ I.\ Shenavrin}
\
\\
\affil{Sternberg State Astronomical Institute,
        Moscow State University, Universitetskij Prospekt 13,
        Moscow, 119899, Russia}

\begin{abstract}
We combine all published NIR  and  optical photometrical observations
of NGC 4151 as well as our new unpublished yet data which can be used for
determination of time delays between optical and NIR variations.
Before we have found change of time delay value for variations
in $K$ and $L$ filters for different states of the luminosity. Here
we are considering new data for deep minimum after the very
high state of the nucleus. We conclude that dust recovering time
after  the high state is for any case more then
several years.

\end{abstract}
\thispagestyle{plain}

\section{Introduction}

 The review of past
 results on the optical-NIR time delay  investigations  for this object
 as well as for other AGNs can be found in Oknyanskij (2002), Oknyankij
and Horne (2001). Here we take into account and discuss new result
for NGC 4151  Minezaki et al. (2004).

In past  papers  (Oknyanskij, 1993, Oknyanskij et al., 1999) has been found
time delay values between optical UBV  and NIR (JHKL)  variations  for two
different luminosity cyrcles of NGC~4151 activities (see Fig.1): Case A
(1969-1980) and Case B (1990-1998).
We have found that time delay  is longer for Case B following
to significantly more high level of luminosity in this cyrcle comparative
to the past case A.

         We  are  going  to  report  about the new time delay  determinations
using  optical  UBV   and   NIR data for NGC 4151  in the minimum after high
 state B (then we call interval 1999-2004 as  Case C).

The past NIR time delays  for NGC~4151  at Cases A and B as well as new
ones for Case C (this work and Minezaki at al. 2004) are collected
in Table 1.

\begin{table}
\caption{Collects optical--infrared time delays measured in NGC~4151 }

\begin{tabular}{lcccccl}
\\ \tableline
object  & delay         & bands &         reference
\\      & (d)           &       &
\\ \tableline
NGC~4151 (Event A)
\\      & $30-60$       &                      & Penston 1974
\\      & $18\pm6$      & $\it K(U)$           & Oknyanskij 1994
\\      & $26\pm6$      & $\it L(U)$           & Oknyasnkij $\&$ Horne 2000
\\
NGC4151 (Event B)
\\      & $\sim6$       & $\it J(UBV)$        & Oknyanskij et al. 1999
\\      & $8\pm4$       & $\it H(UBV)$        &
\\      & $35\pm8$      & $\it K(UBV)$        &
\\      & $97\pm10$     & $\it L(UBV)$        &
\\ 
NGC4151 (Event C)
\\      & $94\pm10$       & $\it H(UBV)$        &  This work
\\      & $48\pm2$      & $\it K(V)$          &  Minezaki at al. 2004
\\      & $104\pm10$      & $\it K(UBV)$       &  This work
\\      & $105\pm10$     & $\it L(UBV)$        &

\\
\\ \tableline
\end{tabular}
\end{table}

\subsection{Data}
Details and needed references about used here optical and NIR data can be
found at  Oknyanskij et al.  (1999). Here we
are using for analysis published in part UBV and NIR data for Case C till 2004.
The details and all references  can be found in our past paper Oknyanskij (1999)
and in   Lyuty (2004), which has to be published soon.

\begin{figure}
\vspace{4cm}
\plotfiddle{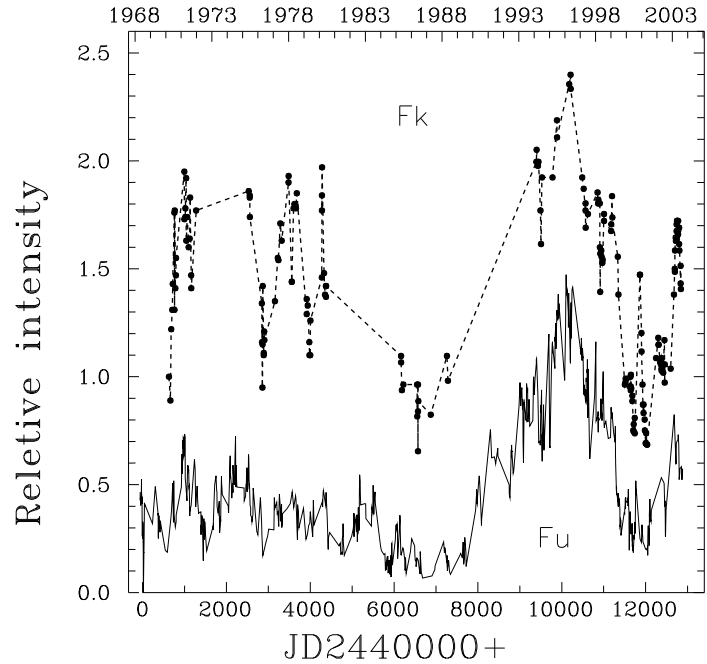} {5 cm}{0}{70}{70}{-185}{-185}
\caption{Overall $K$ light curve (${F_K}=2.512^{9-K}$) reduced to 12" aperture
in comparison with optical $U$ flux variation (${F_U}=2.512^{11-U}$)
through 27" aperture (solid line).
}
\end{figure}

\section{Cross-Correlation analysis}
For cross-correlation analysis we are using the same MCCF method  as
in our past  papers (see details and references in Oknyanskij  1993,
Oknyanskij at al. 1999, Oknyanskij \& Horne, 2001, Oknyanskij 2002).
 The ratio of slow-to-fast
variations amplitudes are much bigger for the Case C than it was before.
Due to the reason  we  need to remove the  slow components in the optical and NIR
light curves before using them for cross-correlation.
This operation does not change the results  but makes the peak in cross-correlation
significanty more sharper.

\begin{figure}
\vspace{4cm}
\plotfiddle{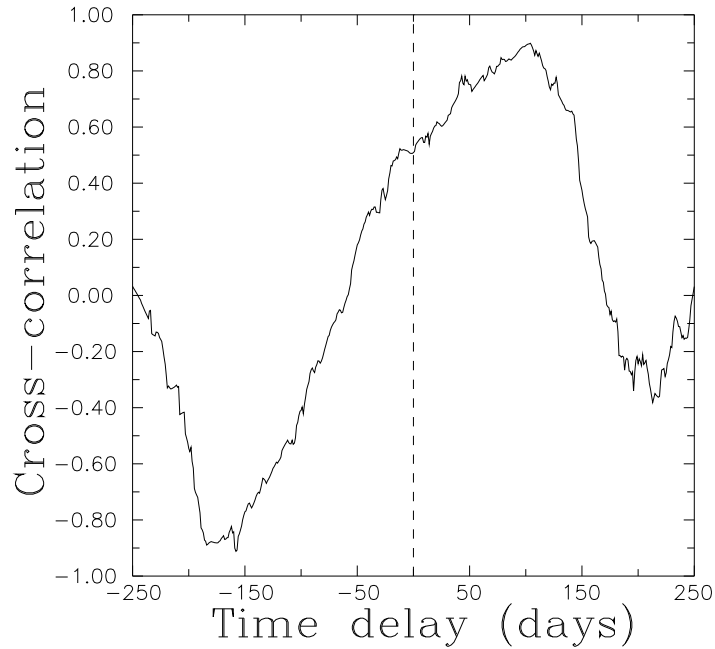} {5 cm}{0}{70}{70}{-185}{-185}
\caption{Cross-correlation function of the optical ($F_B$ ) and  IR ($F_K$)
fast component variations in Case C.  The lag of IR variability
is about 100 days.
}
\end{figure}
As it is seen from the Fig.2  time delay for $K$ variations from optical
ones is about 100 days.

Our data are in very good coincidences with light curves in Minezaki
et al (2004). We were  able to combine the data with our results and
then use the combined data for cross-correlation analysis. We have got
time delay between $V$ and $K$ varations about 50 days for  the same
time interval as in the Minezaki et al. Meanwhile the result has not got
any confirmation in our analysis with all optical data but
 the same time interval data for $K$ data as in Minezaki et al.
Minezaki et al.   Really Minezaki et al have used for analysis just  two
short and smooth curves with only one minimum in them.
If IR and optical data have some difference at additional slow trend then
obtained value for time delay might be very significantly depending from this
fact.

\section{Discussion}

    The dust (graphite grains) can not  be survived on distances
from  nucleus  closer than some critical value. So radiation  of
central source cleans dust from the inner region leaving a  hole
in its distribution.  Radii  of these holes can be estimated using
NIR variations time delays.  These  time  delays probably
can give us redshift-independent luminosities  of AGNs  Oknyanskij
(2000, 2002).
This interpretation has obvious problem if take into account strong
variability of AGNs. If the grains are depleted
when the UV luminosity peaks, and cannot reform, then a dust-free hole
surrounding the central source will be created with radius corresponding
to the sublimation distance at the UV peak (Barvanis, 1992). So to explain
short NIR time delay values in Case A (in despite of known from historical
light curve previous very high states) we have to  involve some explanation  for recovering
or reformation for the dust particles. One of the way has been considered
by Barvanis (1992)  - the dust partical can servive  in  clouds
into the sublimation radius for the peak cases.
The high state  at case B was untypicaly long and dust particls probably
were sublimated very significantly. So it is interesting to find from
observatios how fast the dust can be recavered or reform.

\section{Conclusion}
Time delay between  optical and NIR    variations is a variable
value and it is changed with state of the nucleus  activity.
We have found that time delay for $K$ variations in the last low
state  is about 100 days,  ie  about the same as it  is for $L$ in Cases
B and C.

In a case NGC 4151 we have an opportunity to investigate changes
of  the  dust hole radius  in different activity  states of  the
nucleus and estimate that the time needed for   recovering  or reformation
for the dust particles is  longer than several years.
Alternative explanations can be: anisotropy of radiation field,
shielding  of the central source on a light of sight,  and  also
special orientation of the dust region.

  \end{document}